\documentstyle[12pt]{article}

\newcommand{\be}{\begin{equation}}
\newcommand{\ee}{\end{equation}}
\newcommand{\bea}{\begin{eqnarray}}
\newcommand{\eea}{\end{eqnarray}}
\newcommand{\bean}{\begin{eqnarray*}}
\newcommand{\eean}{\end{eqnarray*}}
\newcommand{\eqn}[1]{(\ref{#1})}

\newcommand{\I}{\mbox{\rm I} \hspace{-0.5em} \mbox{\rm I}\,}
\newcommand{\R}{\mbox{I \hspace{-0.82em} R}}

\newcommand{\Z}{\:\mbox{\sf Z} \hspace{-0.82em} \mbox{\sf Z}\,}
\newcommand{\alg}{{\cal A}}
\newcommand{\del}{\partial}

\def\thebibliography#1{\section*{REFERENCES\markboth
 {REFERENCES}{REFERENCES}}\list
 {[\arabic{enumi}]}{\settowidth\labelwidth{[#1]}\leftmargin\labelwidth
 \advance\leftmargin\labelsep
 \usecounter{enumi}}
 \def\newblock{\hskip .11em plus .33em minus -.07em}
 \sloppy
 \sfcode`\.=1000\relax}

\begin{document}

\begin{flushright}
Napoli: DSF-T-7/94\\
IC/94/76\\
\end{flushright}
\vskip 2.5cm\begin{center}
{\LARGE \bf Distances on a Lattice\\
\ \\
 From Noncommutative Geometry}
\end{center}
\vskip 1.0cm
\centerline {\large G. Bimonte$^{1,2}$,
			F. Lizzi$^{1,3}$,
                    G. Sparano$^{1,3}$
\footnote{\small BIMONTE@ICTP.TRIESTE.IT, LIZZI@NA.INFN.IT, SPARANO@NA.INFN.IT}
}
\vspace{.75cm}
\centerline {\it $^1$ INFN, Sezione di Napoli, Napoli, Italy.}
\vspace{2.25mm}
\centerline {\it $^2$ International Centre for Theoretical Physics,\\
P.O. Box 586, I-34100, Trieste, Italy.}
\vspace{2.25mm}
\centerline {\it $^3$ Dipartimento di Scienze Fisiche, Universit\`a di
Napoli,}
\centerline{\it Mostra d' Oltremare, Pad. 19, I-80125, Napoli, Italy.}

\vspace{.5cm}
\begin{abstract}
Using the tools of noncommutative geometry we calculate the distances between
the points of a lattice on which the usual discretized Dirac operator has been
defined.
We find that these distances do not have the expected behaviour, revealing that
from the metric point of view the lattice does not look at all as a set of
points sitting on the continuum manifold. We thus have an additional
criterion for the choice of the discretization of the Dirac operator.
\end{abstract}

\vfill
\begin{center}
April 1994
\end{center}
\newpage
\setcounter{footnote}{1}

\section{Introduction}

Under the generic name of Non--Commutative Geometry \cite{ConnesBook}
nowadays is meant an approach to the geometry of topological spaces (or
their generalization) based on the algebra of continuous functions with
complex values, or a generalization to a non--abelian algebra. The stress
thus shifts from the topological space as a set of points, to the functions
which can be defined on it.
While the
name of noncommutative geometry should in principle only be given to the
generalizations, it is customary the use of the term even for the case
of topological (Hausdorff) spaces
(see for example \cite{ConnesBook,ConnesLott,JoePepe} and references therein).
 In the
philosophy of Non--Commutative Geometry the main object to study is an
algebra $\alg$, from which (in the abelian case) one can reconstruct an
Hausdorff topological space $M$. It is also possible to introduce algebraic
tools to analyze metric, differential and algebraic topological features.
In particular, with the help of a Hilbert Space on which
the algebra is realized as operator algebra, and of a Dirac Operator, is
then possible to consider also the metric properties of the space, such as
the distances between points.

In this paper we will apply these ideas to the lattice discretizations of
manifolds. The discretization of a manifold, and consequently the
change of
differential equations into finite difference ones, is not always
harmless. It is known \cite{discrebook} that not all solutions of the
discretized
equations in the continuum limit give rise to solutions of the related
differential equations. In the lattice discretization of
field theory the well known undesired feature is the appearance
of spurious states (fermion doubling) (see for example \cite{fermiondoubling}
and references therein).
Here we will concentrate on the metric properties of the lattice, as derived
from the tools of noncommutative geometry\footnote{For a review of approaches
to lattice physics based on noncommutative geometry see \cite{DimakisMuller}
and references therein.} and in particular from the usual discretized Dirac
operator.

We find some surprising results. While in the limit of large separation (as
compared to the lattice spacing) distances behave as expected, for points which
are only a few lattice sites apart, some interesting anomalies appear. In the
case of an hypercubic lattice, we find that
points belonging to a straight line parallel to a coordinate axis ar not
`on a line', in the sense that distances between sites two apart is not twice
the distance between prime neighbours, and in general the distances are not
additive, although they become so in the limit of large separations.

\section{Topological and Metric aspects from Algebras.}

The central idea behind non--commutative geometry is in the work of Gel'fand
and Naimark (see for example \cite{ConnesBook,FellDoran}), which finds a one to
one correspondence between Hausdorff topological spaces and commutative
$C^*$-algebras, that is a commutative algebra $\alg$ on which a norm $\parallel
\cdot \parallel$ and an antilinear involution * are defined, such that
$\parallel f\parallel =\parallel f^*\parallel $, $\parallel f^*f\parallel
=\parallel f^*\parallel ~ \parallel f\parallel $ and $(fg)^*=g^*f^*$ for
$f,g\in \alg $. The algebra $\alg$ is also assumed to be complete in the given
norm.

The correspondence in one sense is simply made by considering the algebra of
complex functions on the topological space. In the other direction the
correspondence is between the algebra and the space of its maximal
ideals\footnote{An equivalent reconstruction can be made by considering the
space of irreducible representations of the Algebra \cite{FellDoran}.} $\hat
\alg$. An ideal $I$ is a subalgebra with the property\footnote{For
noncommutative algebras a distinction between right and left ideals must be
made.}:  $ab\in I, \forall a\in I, b\in\alg$. A maximal ideal is an ideal which
is not contained in any other ideal (apart from the whole algebra $\alg$). The
space $\hat \alg$ is equipped with the topology defined by giving the closure
of any given set of ideals $W$. The closure is defined as follows
\be
\bar W = \{I \in \hat \alg: I \supset \bigcap J, J\in W \}\ \ .
\ee

Once the topological space has been reconstructed, one can see that the maximal
ideals correspond to the functions vanishing in one point, and it is not
difficult to see that the topology above defined corresponds to the topology of
the space with respect to which the functions are continuous,

The power of the method is its ability to be generalizable to noncommutative
$C^*$-algebras. In this case one cannot in general talk of a structure space,
although it is still possible in some cases to reconstruct a topological space,
which however will not be a Hausdorff separable space \cite{poset2}.

A fundamental ingredient in Connes' noncommutative geometry is a $k$-cycle, it
contains in fact the information about the metric. A k-cycle is a triple
$(\alg,H,D)$ consisting of an algebra $\alg$, an Hilbert space $H$ on which the
elements of $\alg$ are realized as linear operators, and a Dirac operator $D$
which is a selfadjoint operator with compact resolvent such that, for a dense
subset of $\alg$, the commutator $[D,f]$ is a bounded operator.  For the
purpose of this paper, what we are interested in is mainly the distance, which
can be reconstructed from the $k$-cycle in the following way: the distance
$d(x,y)$ between two points $x$ and $y$ is defined as
\be
d(x,y) = \sup_f \{|f(x) - f(y)| : f \in \alg, \parallel[D,f]\parallel\leq
1\}.\label{dist1} \ \ ,
\ee
where the norm among operators on $H$ is the one given by the supremum over
normalized vectors of $H$.
This distance is to be compared with the usual definition of (geodesic)
distance
for a Riemannian manifold with euclidean metric $g_{\mu \nu}$:
\be
\bar d(x,y) = \inf_\gamma l_\gamma(x,y)\label{dist2}
\ee
$l_\gamma(x,y)$ being the length of the path $\gamma$ from $x$ to $y$.

To see the relation between these two definitions of distance, consider
the usual dirac operator $D= i\gamma^\mu ( \partial_\mu + A_{\mu})$
acting on some dense
domain of the Hilbert space $H$ of square integrable spinors $\Psi(x)$
defined in a $\R^4$ ($A_{\mu}$ is the connection for some gauge group $G$
and the spinors belong to some representation of $G$). Let
$\alg$ be the algebra of continuous functions of
$\R^4$ vanishing at infinity, and consider them as operators on $H$ acting
by pointwise multiplication. The commutator $[D,f]$, acting on the spinor
$\Psi$, is
\bea
[D,f]\Psi &=& (i\gamma^\mu\partial_\mu f)\Psi + if\gamma^\mu\partial_\mu\Psi -
if\gamma^\mu\partial_\mu\Psi \\
          &=& i(\gamma^\mu\partial_\mu f)\Psi.
\eea
The norm of the commutator is thus the maximum of $\sqrt{\del^\mu f^* \del_\mu
f}$, which is
also equal \cite{JoePepe} to the
Lipschitz norm of $f$ defined as \be
\parallel f\parallel_{Lip}:= \sup_{x\neq y} {|f(x) - f(y)|\over \bar
d(x,y)}\ \ .
\ee
Thus
\be
\parallel[D,f]\parallel = \max \sqrt{\del^\mu f^* \del_\mu
f}= \parallel
f\parallel_{Lip}:= \sup_{x\neq y} {|f(x) - f(y)|\over \bar
d(x,y)}\label{Lip} \ee
To check that the definitions of distance in \eqn{dist1} and \eqn{dist2}
agree, consider that the condition on the norm of $[D,f]$ in \eqn{dist1},
and equation \eqn{Lip} imply
\be
d(x,y)\leq\bar d(x,y) \ \ .
\ee
Considering now $\bar f(t)=\bar d(x,t)$ as a function of $t$ (suitably
regularized), it satisfies
\be
\parallel[D,\bar f(t)]\parallel = \parallel\bar d(x,t)\parallel_{Lip} = 1
\ee
which gives
\be
d(x,y) = \bar d(x,y).
\ee

Once the notion of distance between pairs of points is given, the other
metric properties can be obtained with the traditional techniques.

Thus, we have seen that, in order to reconstruct the topological and metric
aspects of a topological space one should be given an abelian algebra
$\alg$ realized as linear operators on a Hilbert space $H$, together with
another operator, the Dirac operator $D$.

\section{Abelian Discretization of a Manifold}

The discretization of a continuous manifold via a regular lattice is a well
known technique, which in general provides some simplifications of the
problem at hand, such as the substitution of differential equations with
finite difference equations, or the possibility to encode with a finite set
of numbers the state of a system. Invariably something is missing when a
continuous manifold is approximated by a lattice. A prime victim is
topology, the points of a lattice are equipped with the discrete topology,
each single point is both open and closed, and every map is a
continuous function, that is, the inverse image of an open set is open.
This for the simple reason that every collection of points of a lattice is
open. The programme of \cite{Sorkin,poset1} is in fact to define
generalizations of lattices with non trivial topology. In \cite{poset2} we
have discussed as a lattice with non trivial topology is, in some sense, a
non--commutative manifold. In this paper we will however limit ourselves to
the customary, `abelian', lattices.

In order to discretize a manifold one chooses a finite or countable set of
points, so that the state of the system, which in general will be a function on
the manifold, is now the finite or countable set of values of the function on
the points of the lattice. Similarly, differential operators are approximated
by finite differences ones. The hope is that even a limited number of points
will capture the essential features of the continuous system. Obviously the
discretized system cannot have exactly the same behaviour of the continuous
one. The idea is that by increasing the number of points, in other words by
taking the continuum limit of the system, the correspondence becomes exact.

{}From the noncommutative geometry point of view,
for a $N$ point discretization, the algebra consists of ordered $N$-ples of
complex numbers, with product and sum acting component by component. The number
$N$ can be taken to be
infinity without troubles.
{}From this algebra we reconstruct the abelian lattice, that is a set
of points with the discrete topology, each point is both closed and open,
and therefore so is any subset. On the lattice we then consider the customary
discrete version of Dirac operator. We then calculate the distances between the
points of the lattice, with the methods of noncommutative geometry.

Let us consider in detail the case of a four-dimensional euclidean
space-time $\R^4$ (with the usual metric).
A possible discretization of the manifold $\R^4$ is then accomplished by
considering an hypercubic lattice
of points with integer coordinates (in units of the lattice spacing
$a$):
\be
x \equiv a (i_1, i_2, i_3, i_4)\ \ , i_{\mu}\in\Z \ \ , \label{lattice}
\ee
together with the Hilbert space of ``square integrable" spinors:
\be
\Psi_x:\
\sum_{x}\Psi_x^{\dagger}\Psi_x =
\sum_{x,\alpha,i}\Psi_{i \alpha x}^*\Psi_{i \alpha x}
< \infty \ \ .
\ee
where $i$ and $\alpha$ denote spinor and gauge indices respectively.
This is the Hilbert space commonly called $l^2$. On this space the elements of
the algebra $f=\{f_x\}$ are realized (as operators) as diagonal matrices (we
use the same symbols for the operator and the element of the algebra):
\be
f_{x,i,\alpha;y,j,\beta}=f_x\delta_{xy}\delta_{\alpha \beta}\delta_{ij}\
\mbox{\rm no sum on}~ x. \ee

Given this discretization the next question is the choice of the finite
difference operator
which corresponds to the Dirac operator. This choice is not unique:
the simplest one is that made by Wilson:
\be
(D \Psi)_x= \sum_{\mu} \gamma^{\mu} (D_{\mu} \Psi)_x~,
\label{customary} \ee
where $D_{\mu}$ is the covariant discrete derivative in the direction
$\mu$:
\be
(D_{\mu}\Psi)_x={1\over 2 i a} (U^{\mu}_x \Psi_{x+\mu} -
U^{-\mu}_{x-\mu}\Psi_{x-\mu}) \ \ \label{covariant}.
\ee
In the latter equation $x+\mu$ denotes the site displaced from $x$ by one
unit along direction
$\mu$ and $U^{\mu}_x$ is the gauge group element (concretely a unitary
matrix) associated with the link
joining $x$ with $x+\mu$, the so-called link-variable. By definition,
$U^{-\mu}_x=(U^{\mu}_x)^{-1}$.

Despite its simple look, this Dirac operator hides subtle problems. As is
well
known it is affected by the pathology of ``species doubling": in the
continuum limit, $a\to 0$, it describes 16 Dirac fermions with degenerate
masses. For the time being we will not worry about this question and will
come back to it at the end of section 5.
We will now show instead that at the even more fundamental level of the
definition of distances, an important pathology happens.

\section{Distances on the Hypercubic lattice}

In this section, we will compute the distance between the points
of the lattice \eqn{lattice}, using as Dirac operator the finite
difference operator \eqn{customary}. For simplicity, we will limit
ourselves to the distance $d_k$ between the origin of the lattice and a point
$x=(k,0,0,0)$ of the $x-$axis (it will be clear from the proof that the
distances are invariant under translations and rotations that leave the
lattice invariant).
Naively, one would
expect to find for $d_k$
the following result:
\be
d_k=a|k| \ \ . \label{naive}
\ee
As we shall see in the sequel, this expectation is incorrect.

The computation of $d_k$ consists of two parts: first, we show how to
reduce the problem to that of finding the distances on a one-dimensional
lattice endowed with a ``Dirac" operator given by the
flat discrete derivative in one dimension, namely eqn.
\eqn{covariant}
with $\mu=1$ and link-variables all equal to $\I$:
\be
(D_1^{(0)} \Psi)_x=\frac{1}{2ia}(\Psi_{x+1}-\Psi_{x-1})~.\label{flat}
\ee
We then proceed
in the next section with the
computation of the distances for this simpler one-dimensional lattice.

In order to achieve the reduction to a one-dimensional problem, we start
by proving that the supremum of (2) can be attained considering functions
which depend only on the $x-$ coordinate of the lattice sites. In fact, let
$f=f(i_1,i_2,i_3,i_4)$ be any function saturating eq.(2):
$$
\parallel [D,f] \parallel \le 1~,
$$
\be
d_k=|f(k,0,0,0)-f(0,0,0,0)|~.
\ee
Consider now the function $f^{(x)}$ defined as:
\be
f^{(x)}(i_1,i_2,i_3,i_4)=f(i_1,0,0,0)~.
\ee
That is we consider a function which coincides with $f$ on the $x_1$ axis, and
is constant along the other three axis.
By construction $f^{(x)}$ depends only on $i_1$ and gives for $d_k$ the same
result as $f$. Then, if we show that $\parallel [D,f^{(x)}]\parallel
 \le 1$,
the statement will be proven. Now, for any normalized $\Psi \in l^2$, we
have:
\be
\parallel [D,f^{(x)}]\Psi\parallel ^2=
\parallel \gamma^1[D_1,f^{(x)}]\Psi\parallel^2=
\parallel [D_1,f^{(x)}]\Psi\parallel^2~.
\ee
The norm
$\parallel [D_1,f^{(x)}] \parallel $ is easily seen to be
independent on the link variables.
Consider in fact the following gauge transformation $\cal U$:
\be
\hat \Psi_x=({\cal U} \Psi)_x=g_x\Psi_x
\ee
where $g_x$ are unitary matrices such that:
$$
g_{0,i_2,i_3,i_4}= g_{1,i_2,i_3,i_4}= \I~,
$$
$$
U_x^{1}g_{x+1}=U^{-1}_{x-1}g_{x-1}~~~~~~~~\rm{for} ~i_1>0
$$
\be
U_{x-1}^{-1}g_{x-1}=U^{1}_{x}g_{x+1}~~~~~~~~\rm{for}~ i_1<1~.\label{gauge}
\ee
It is easy to verify that:
$$
\parallel [D_1,f^{(x)}] \parallel\equiv
\sup_{\parallel \Psi \parallel^2=1}
\parallel [D_1,f^{(x)}]\Psi \parallel =
\sup_{\parallel \Psi \parallel^2=1}
\parallel [D_1,f^{(x)}] {\cal U} \Psi \parallel =$$
\be
\sup_{\parallel \Psi \parallel^2=1}
\parallel [
D^{(0)}_1,f^{(x)}] \Psi \parallel =
\parallel [D^{(0)}_1
,f^{(x)}] \parallel~, \label{norm1}
\ee
Now, the Hilbert space $l^2$ can be decomposed into a direct sum of
subspaces left invariant by the operator $[D^{(0)}_1,f^{(x)}]$:
\be
l^2= \bigoplus _{l,m,n;j}
H^{(l,m,n;j)}~,
\ee
where
\be
H^{(l,m,n;j)}\equiv \{\Psi: \Psi_{i,\alpha,x}=0 ~{\rm unless}~
i_2=l,~i_3=m,~i_4=n~ {\rm and}~ i=j\}~.
\ee
Moreover, the restrictions of $[D^{(0)}_1,f^{(x)}]$ to these subspaces are
all equal to each other and then we can write:
$$
\parallel [D^{(0)}_1,f^{(x)}] \parallel =
\sup_{(l,m,n;j)}\{
\parallel [D^{(0)}_1,f^{(x)}] \parallel _{(l,m,n;j)}\}~=
$$
\be
=\parallel [D^{(0)}_1,f^{(x)}] \parallel _{(0,0,0;1)}~=
\parallel [D_1,f^{(x)}] \parallel _{(0,0,0;1)}~, \label{norm2}
\ee
where
$\parallel [D_1,f^{(x)}] \parallel _{(l,m,n;j)}$
denotes the
norm of the restriction of $[D_1,f^{(x)}]$ to $H^{(l,m,n;j)}$.

But it is now easy to check that, for any normalized
$\Psi \in H^{(0,0,0;1)}$, we have:
\be
\parallel [D_1,f^{(x)}]\Psi \parallel \le
\parallel [D,f]\Psi \parallel \le 1~.
\ee
and then, in view of \eqn{norm1} and \eqn{norm2}, it follows that:
\be
\parallel [D_1,f^{(x)}]\parallel =
\parallel [D^{(0)}_1,f^{(x)}]\parallel_{(0,0,0;1)}
\le 1~.
\ee
What this equation proves is that the distances $d_k$ coincide with those
for a one-dimensional lattice endowed with the flat-derivative operator of
eqn.
\eqn{flat}, regarded as Dirac operator. We have thus reduced the calculation of
the distances on a one-dimensional lattice. We calculate these distances in the
next section.

\section{ Distances on the one-dimensional lattice}

In this section, we compute the distances on a one-dimensional lattice:
\be
x_k=ak~~,k\in\Z~.
\ee
with the Dirac operator given by eqn. \eqn{flat}. For economy of notation
this
operator will be denoted in this section by the letter $D$.
First of all, it is clear that distances are translationally invariant. It
is enough then to compute the distance $d_k$ between sites $x_0$ and
$x_k$. Our proof goes as follows: in order to obtain the supremum in
\eqn{dist1} we first show that this can be attained considering only real
`step functions', functions which are constant except for a finite set of
points, on which they monotonically increase. The distance between the
points is then simply the total step. With these functions the
problem becomes the one of finding the maximum step the function can have,
subject to the restriction that the norm of the commutator is less than
one. For step functions this last requirement
will mean that the eigenvalues of a finite matrix
are all less than one in modulus.

As a first step, we will prove the distance $d_k$ can be obtained by
taking the supremum of (2) over the set of real functions
$f^{(k)} \equiv \{f^{(k)}_i;~i \in \Z\}$ such that:
\begin{itemize}
\item[$a)$]
$f^{(k)}_i=f^{(k)}_0 ~\forall i < 0~,~ f^{(k)}_i=f^{(k)}_k
{}~\forall i > k, $
\item[$b)$]
$f^{(k)}_j \le f^{(k)}_i ~\forall j<i,$
\item[$c)$]$\parallel [D,f^{(k)}]\parallel \le 1~.$
\end{itemize}
In fact, let
$f \equiv \{f_j~;~j \in \Z\}$ be an arbitrary complex function
on the lattice. We then have:
\be
([D,f] \Psi)_j= \frac{i}{2a}[(f_{j+1}-f_j) \Psi_{j+1} +
(f_j -f_{j-1})\Psi_{j-1}).
\ee
In what follows we will set $2a=1$, the changes for a general $a$ being
trivial. For any $\Psi \in l_2$, consider
a new vector $\hat {\Psi}$ such that:
$$
\hat {\Psi}_j=e^{i \phi_j}\Psi_j ,
$$
\be
\arg((f_{j+1}-f_j)\hat{\Psi}_{j+1})=\arg((f_j-f_{j-1})\hat{\Psi}_{j-1}),
{}~\forall j \in \Z. \label{arg}
\ee
Such a vector always exists (for example, one can choose $\hat {\Psi}_0=
\Psi_0$ and $\hat {\Psi}_1=
\Psi_1$ and use eq. \eqn{arg} to recursively determine $\hat{\Psi}_j$ for the
other  $j$'s). We then have:
\bea
\parallel [D,f]\Psi\parallel^2 &=& \sum_j |(f_{j+1}-f_j) \Psi_{j+1} +
(f_j-f_{j-1}) \Psi_{j-1}|^2 \le \nonumber\\
&\le& \sum_j (|f_{j+1}-f_j||\hat{ \Psi}_{j+1}| + |f_j-f_{j-1}|
|\hat{\Psi}_{j-1}|)^2=\parallel [D,f]\hat{\Psi}\parallel^2.
\eea
It follows that
\be
\parallel [D,f] \parallel ^2 = \sup_{c_j \in \R^+, \sum_j c_j^2=1}
\sum_j(|f_{j+1}-f_j|c_{j+1}+|f_j-f_{j+1}|c_{j-1})^2\ \ .
\label{norm}
\ee
Consider now the real, monotonic function $F$ defined by:
\be
F_{j+1}-F_j=|f_{j+1}-f_j|~~~,~~F_0=0.
\ee
Considerations similar to those applied to $f$ imply that the norm of
$[D,F]$ is again given by equation \eqn{norm}. It follows then that
\be
\parallel [D,f] \parallel = \parallel [D,F] \parallel.
\ee
Moreover, we clearly have
\be
|f_k-f_0|=
|\sum_{m=0}^{k-1} (f_{m+1}-f_m)| \le
\sum_{m=0}^{k-1} |f_{m+1}-f_m| = F_k - F_0.
\ee
Finally, consider the function $f^{(k)}$ which coincides with $F$ for
$0 \le i \le k$. Use of equation \eqn{norm}implies that
\be
\parallel [D,f^{(k)}] \parallel \le \parallel [D,F] \parallel.
\ee
{}From the last inequalities it then follows that in order to compute the
distance between any two points we can restrict the supremum to the functions
$f^{(k)}$:
\be
d_k = \sup_{f^{(k)}} \{f^{(k)}_k - f^{(k)}_0 : ~~
\parallel[D,f^{(k)}]\parallel < 1\}\ \ . \label{dk}
\ee

Making use of \eqn{flat},
it is now easy to check that the norm of $[D,f^{(k)}]$ is equal to the
maximum eigenvalue $r$
of the square, symmetric and real $(k+1) \times
(k+1)$ matrix $H$:
$$
H_{m,n}=\Delta_{m} \delta_{m+1,n} +
\Delta_{m-1}\delta_{m-1,n}~~~m=2,\cdots,k;~~n=1,\cdots,k+1.
$$
\be
H_{1,n}= \Delta_1 \delta_{n,2}~~;~~ H_{k+1,n}=\Delta_{k} \delta_{n,k}~~,
\ee
where $\Delta_l = f^{(k)}_{l}-f^{(k)}_{l-1},~l=1,\cdots,k$.
Then equation \eqn{dk} for the distance can be rewritten as:
\be
d_k = \sup_{f^{(k)}} \left\{ \sum_{i=1}^k \Delta_i :  r < 1\right\}.
\ee
Now, according to a theorem of linear algebra on matrices with non
negative elements \cite{matrixbook}, the maximum eigenvalue $r$ of $H$ is
less than 1 if and only if all the leading principal minors
$H_n(\Delta_1,\cdots,\Delta_{n-1}),~n=1,\cdots,k+1$ of the matrix
$\I-H$ are
positive:
\be
H_1=1 >0~~; H_2=1-\Delta_1^2>0~~;\cdots H_{k+1}=\det(\I-H)>0.
\label{inequalities}
\ee
We notice the following recursive relation among the $H_n$:
\be
H_{n+1}=H_n -\Delta^2_n H_{n-1}~,~1<n<k+1\ \ .
\ee
The determination of the distances is thus reduced to finding the
supremum of $\sum_{i=1}^k \Delta_i$ in the open region $X$ of $\R ^{k}$
specified
by the set \eqn{inequalities} of inequalities.
It is clear now that $\sum_{i=1}^k \Delta_i$
cannot have a local maximum in the interior of $X$. It then follows that
the supremum must be reached at some point of the frontier $\partial X$
of $X$, which is the set of points $y$ such
that:\begin{enumerate}
\item At least one among $H_n(y)$ is equal to 0.
\item There is a sequence of points of $X$ converging to $y$.
\end{enumerate}
Now, $\partial X= \bigcup_{i=2}^{k+1} \partial X_i$, where $\partial X_i$
is the open surface given by the equation\footnote{On $\partial X_i$,
the recursion relations
imply $H_l=0$, for $l>i$, if $H_i=0$.}:
\be
\partial X_i=\{(\Delta_1,\cdots,\Delta_k): \Delta_m \ge 0;~H_j>0,~j=1,\cdots
i-1;~H_i=0 \}.
\ee

Then $d_k=\max\{\lambda_i,~i=2,\cdots,k+1\}$, where $\lambda_i$ is the
maximum, when it exists, of $\sum_{j=1}^k \Delta_j$ on $\partial X_i$. We
can now prove that, for $i<k+1$, it must be $\lambda_i=d_{i-1}+d_{k-i}$. In
fact, let $(\Delta^{(i)}_1,\cdots,\Delta^{(i)}_k)$ be the point
of $\partial X_i$ such that $\max (\Delta_1+\cdots +\Delta_k)=(
\Delta^{(i)}_1+\cdots +\Delta^{(i)}_k)=\lambda_i$. Then, it follows
from
the the recursion relation that $ \Delta^{(i)}_{i}=0 $. In fact, if
$ \Delta^{(i)}_i \neq 0$, we would have $H_{i+1}=H_i-
\Delta^{(i)}_i H_{i-1}<0$ and then $(\Delta^{(i)}_1,\cdots,\Delta^{(i)}_k)
\notin \partial X$. But $\Delta^{(i)}_i=0$ implies that the matrix $H$
block diagonal and this obviously implies the statement. Then, if the
distances $d_i,~i=1,\cdots,k-1$ are known we need only compute
$\lambda_{k+1}$. The latter can be found with the method of the Lagrange
multipliers: one looks for the solutions of the following system of
algebraic equations in the unknowns $(\Delta^{(k+1)}_1,\cdots,
\Delta^{(k+1)}_k)$ and $\mu$:
\bea
\left. \frac{\partial}{\partial \Delta_j}\left[ \sum_{i=1}^k \Delta_i +
\frac{1}{2
\mu} H_{k+1} \right ]\right|_{\Delta_j=\Delta^{(k+1)}_j} &=&0 \ \ , \nonumber\\
H_{k+1}&=&0.
\eea
(a posteriori one has to check that $H_i(\Delta^{(k+1)}_1,\cdots,
\Delta^{(k+1)}_{i-1})>0,~i=2,\cdots, k$.). The first few distances can easily
be computed analytically. For an arbitrary $a$ one gets:
\bea
d_1&=&2a\nonumber\\
d_2&=&2 \sqrt 2 a\nonumber\\
d_3&=&4a\nonumber\\
d_4&=&2\sqrt 6 a.
\eea
The distances $d_k$, for $k>4$, have been determined numerically. In the
table
\begin{table}
\begin{tabular}{|l||ccccccccccccc|}
\hline
{\small Sites Apart}&  1 & 2    & 3 &     4 & 5 & 6    & 7 &    8 &  9 & 10
& 11 & 12    & 13 \\
{\small Distance}   &  2 & 2.83 & 4 &  4.90 & 6 & 6.93 & 8 & 8.94 & 10 & 10.95
& 12 & 12.96 & 14 \\
\hline
\end{tabular}
\caption{\sl The distances calculated analytically for the site differences
less than 4, numerically otherwise.}
\end{table}
the distances for sites up to 13 apart are shown.
The results
show that the sites with $k$ even behave differently from those  with $k$
odd. The result can be expressed by the following formulae:
\bea
d_{2l+1}&=&2(l+1)a \nonumber\\
d_{2l}&=&(2l+1)a - \epsilon_{2l}
\eea
where $\epsilon_{2l}>0$. We thus see that there is an overall constant
"anomaly" equal to $a$ for all sites, plus an extra one, $\epsilon_{2l}$,
for the even sites, which vanishes for $2l \rightarrow \infty$. The
conclusion of this analysis is that, form the point of view of the
finite difference operator \eqn{customary}, the lattice does not look at all
as a subset of equidistant points of the real line. Preliminary results seem to
indicate that in order to get the desired distances, one is forced to use
non-local difference operators.

As pointed out at the end of section 3, the `naive" discrete Dirac operator
\eqn{customary} is plagued with the problem of species doubling: in the
continuum limit it describes 16 Dirac fermions with equal masses. Many
possible remedies to this difficulty have been considered in the literature
and one may wonder whether they will cure the pathology in the distances as
well.
We shall see immediately that this is not the case. It will be sufficient
to consider, as an example, the modification of the discrete Dirac
operator introduced by Wilson: one adds to \eqn{customary} a term
proportional to the discretized covariant laplacian, whose effect is
to make the masses of 15 out of the 16 fermions of the order of the
inverse of the lattice spacing. In this way the spurious fermions
disappear in the continuum limit. The modified operator $D_W$ reads :
\be
(D_W\Psi)_x= (D \Psi)_x + \frac{r}{2a}\sum_{\mu}(
U^{\mu}_x\Psi_{x+\mu}+
U^{-\mu}_{x-\mu}\Psi_{x-\mu} -2 \Psi_x)~.
\label{wilson}
\ee
where $r$ is an arbitrary non vanishing real constant. If we take this as
the Dirac operator to be used in the computation of the distances instead
of \eqn{customary}, it is easy to verify that the only change in the
distances will be an overall rescaling by a factor $(1+r^2)^{-1/2}$
(this shows that from the distances point of view the inclusion of the
Laplacian actually worsens the situation instead of improving it, because
now we
do not recover the correct distances even for large separations). The fact
is that, as is clearly shown by the computations of section 3 and 4, the
distances mainly depend on the choice of the discrete flat derivative in
one dimension.

\section{Conclusions}

The first conclusion we can draw is that the Dirac operator usually used on the
lattice does not reproduce the distances one would expect. This
opens another problem. We have three concepts of distance at hand: the naive
one of equation \eqn{naive}; the one we calculated in sections 4 and 5,
which however
in the limit of large separations goes to the naive one; finally we can
envisage some sort of `geodesic' distance:
\be
\tilde d(x_k,x_l)=\sum_{i=k+1}^l d(x_i,x_{i-1}) \ \ .
\ee
This last distance in our case is twice the naive distance for all separations
independently of the lattice spacing $a$, and thus has a different continuum
limit.

One would of course like that all those distances be the same, as in the
continuum case. Since the only ingredient of this calculation is the choice of
the discretization of Dirac operator, we think that we are seeing another
pathology of the choice \eqn{customary}. We do not know of a discretization of
the Dirac operator which reproduces correctly the distances.

In conclusion, the results of this paper show that the concepts of
noncommutative geometry, apart from representing a fashinating generalization
of known geometrical notions, may also suggest new points of view about
conventional problems, like the discretization of a field theory on a
lattice. We have seen that if, on a lattice, one has assigned distances
(inherited, say, from the continuum), noncommutative geometry provides
additional, stringent criteria for the choice of a discretization of the
Dirac operator.

\ \\
{\large \bf Acknowledgements} \\
We would like to thank A.P.~Balachandran,
E.~Ercolessi, G.~Landi and P. Teotonio-Sobrinho for discussions and a useful
correspondence.

\def\up#1{\leavevmode \raise.16ex\hbox{#1}}
\newcommand{\npb}[3]{{\sl Nucl. Phys. }{\bf B#1} \up(19#2\up) #3}
\newcommand{\plb}[3]{{\sl Phys. Lett. }{\bf #1B} \up(19#2\up) #3}
\newcommand{\revmp}[3]{{\sl Rev. Mod. Phys. }{\bf #1} \up(19#2\up) #3}
\newcommand{\sovj}[3]{{\sl Sov. J. Nucl. Phys. }{\bf #1} \up(19#2\up) #3}
\newcommand{\jetp}[3]{{\sl Sov. Phys. JETP }{\bf #1} \up(19#2\up) #3}
\newcommand{\rmp}[3]{{\sl Rev. Mod. Phys. }{\bf #1} \up(19#2\up) #3}
\newcommand{\prd}[3]{{\sl Phys. Rev. }{\bf D#1} \up(19#2\up) #3}
\newcommand{\ijmpa}[3]{{\sl Int. J. Mod. Phys. }{\bf A#1} \up(19#2\up) #3}
\newcommand{\prl}[3]{{\sl Phys. Rev. Lett. }{\bf #1} \up(19#2\up) #3}
\newcommand{\physrep}[3]{{\sl Phys. Rep. }{\bf #1} \up(19#2\up) #3}
\newcommand{\journal}[4]{{\sl #1 }{\bf #2} \up(19#3\up) #4}

%
%
%

\end{document}